\documentclass[11pt,a4paper]{article}
\usepackage[hyperref]{acl2021}
\usepackage{times}
\usepackage{latexsym}
\usepackage{import}
\usepackage{graphicx}
\usepackage{url}
\usepackage{pdflscape}
\usepackage{booktabs, longtable, tabularx,makecell}
\usepackage{rotating}
\usepackage{multirow}
\usepackage{xspace}

\usepackage{enumitem}
\usepackage{booktabs}
\usepackage{xcolor}
\usepackage{soul}
\usepackage{amssymb}

\usepackage{microtype}

\aclfinalcopy

\newcommand{\cmmnt}[1]{}

\newcommand{\secref}[1]{\S\ref{#1}\xspace}

\newcolumntype{P}[1]{>{\centering\arraybackslash}p{#1}}
\newcolumntype{L}{>{\RaggedRight\arraybackslash\hspace{0pt}}X}
\title{Code to Comment Translation: A Comparative Study \\ on Model Effectiveness \& Errors}

\author{Junayed Mahmud, Fahim Faisal, Raihan Islam Arnob,\\ \textbf{Antonios Anastasopoulos, Kevin Moran} \\
  Department of Computer Science\\
  George Mason University, USA  \\
  \texttt{jmahmud,ffaisal,rarnob,antonis,kpmoran@gmu.edu}}
  
\date{}

\begin{document}
\maketitle

\begin{abstract}
Automated source code summarization is a popular software engineering research topic wherein machine translation models are employed to ``translate'' code snippets into relevant natural language descriptions. %that has evolved from using template based-approaches to more recent neural models, which have proven to be reasonably capable. 
%To evaluate code machine translation-based metrics are typically used. 
Most evaluations of such models are conducted using automatic reference-based metrics. However, given the relatively large semantic gap between programming languages and natural language, we argue that this line of research would benefit from a qualitative investigation into the various error modes of current state-of-the-art models.    %However, to gain a more holistic understanding of code summarization models there is a need to qualitatively investigate the errors such models make in the course of their translations in order to grasp their practical utility. 
Therefore, in this work, we perform both a quantitative and qualitative comparison of three recently proposed source code summarization models. In our quantitative evaluation, we compare the models based on the smoothed BLEU-4, METEOR, and ROUGE-L machine translation metrics, and in our qualitative evaluation, we perform a manual open-coding of the most common errors committed by the models when compared to ground truth captions. Our investigation reveals new insights into the relationship between metric-based performance and model prediction errors grounded in an empirically derived error taxonomy that can be used to drive future research efforts.\footnote{Our annotations and guidelines are publicly available on Github \url{https://github.com/SageSELab/CodeSumStudy} and Zenodo: \url{https://doi.org/10.5281/zenodo.4904024}.} %Our qualitative evaluation illustrates that the CodeBERT model achieves the highest performance with a smoothed BLEU-4 score of 24.15. Moreover, we performed an empirical study by comparing the predictions generated by three deep learning models with the original descriptions. A taxonomy is proposed based on the results obtained from the study.
\end{abstract}

\section{Introduction and Motivation}
\label{sec:intro}

Proper documentation is an important component of modern software development, and previous studies have illustrated its advantages for tasks ranging from program comprehension~\cite{Garousi:IST'15} to software maintenance~\cite{Chen:JSS'09}. However, manually documenting software is a tedious task~\cite{automaticDocumentation} and modern agile development practices tend to champion working code over extensive documentation~\cite{beck2001agile}. As such, a range of important documentation activities are often neglected~\cite{Zhi:JSS'15} leading to deficiencies in carrying out development activities and contributing to technical debt. Because of this, researchers have worked to develop automated code summarization techniques wherein machine translation models are employed to generate precise, semantically accurate natural language descriptions of source code~\cite{programComprehension}. Due to the promise and potential benefits of effective automated source code summarization techniques, this area of work has seen constant and growing attention at the intersection of the software engineering and natural language processing research communities~\cite{codeSummarizationSurvey}. 

Various techniques for automated source code summarization have been explored extensively over the past decade. Some of the earliest approaches made use of a combination of structural code information and text retrieval techniques for determining the most relevant terms~\cite{programComprehension}, with follow up work investigating the use of topic modeling~\cite{evaluateSummarization}. Techniques then evolved from using information retrieval to canonical machine learning techniques, with~\citet{codeFragment} using supervised Naive Bayes and Support Vector Machine classifiers to identify code fragment lines that could be used as suitable summaries.  One of the first appearances of language modeling came from~\citet{compareComments} who proposed an approach combining a software word usage model, natural language generation systems, and the PageRank algorithm~\cite{pagerank} to generate summaries. Driven by the advent of deep learning, current state-of-the-art techniques generally make use of large-scale neural models and have significantly improved the performance of code summarization tasks. For instance, \citet{iyer_summarizing} used Long Short Term Memory~\cite{lstm1997} with attention~\cite{Bahdanau2015} to generate summaries from a code snippet. Following this work, researchers have applied several deep learning-based approaches to the task of source code summarization~\cite{retrievalBased,reinforcementLearningBased,GNNBased}. 

In most works on automated code summarization, the performance of the generated natural language descriptions is evaluated using reference-based metrics adapted from machine translation, e.g., BLEU~\cite{BLEU} and METEOR~\cite{meteor}, or text summarization, e.g., ROUGE~\cite{rouge}. As such, most researchers make conclusions based on the results obtained using these metrics. However, the code summarization task is a difficult one -- due in large part to the sizeable semantic gap between the modalities of source code and natural language. As such, while these metrics provide a general illustration of model efficacy, it can be difficult to determine the specific shortcomings of neural code summarization techniques without a more extensive qualitative investigation into their errors.

Few past studies have examined the failure modes of neural code summarization models as we outline in \secref{sec:related}. Therefore, to further explore this topic, in this paper we perform both a qualitative and quantitative empirical comparison of three neural code summarization models.  Our quantitative evaluation offers a comparison of three recently proposed models (CodeBERT~\cite{codebert}, NeuralCodeSum~\cite{neuralCodeSum}, and code2seq~\cite{code2Seq}) on the Funcom dataset~\cite{funcomPaper} using the smoothed BLEU-4~\cite{lin2004orange}, METEOR~\cite{meteor}, and ROUGE-L~\cite{rouge} metrics whereas our qualitative evaluation consists of a rigorous manual categorization of model errors (compared to ground truth captions) based on a procedure adapted from the practice of open coding~\cite{openCodingBook}. In summary, this paper makes the following contributions:  

\begin{itemize}[noitemsep,leftmargin=*,nolistsep]
    \item We offer a quantitative comparative analysis of the CodeBERT, NeuralCodeSum, and code2seq models applied to the task of Java method summarization in the Funcom dataset. The results of this analysis illustrate that the CodeBERT model performs best to a statistically significant degree, achieving a BLEU-4 score of 24.15, a METEOR score of 30.34, and a ROUGE-L score of 35.65.
    \item We conduct a qualitative investigation into the various prediction errors made by our three studied models and derive a taxonomy of error modes across the various models. We also offer a discussion about differences in errors made across models and suggestions for model improvements.
    \item We offer resources on GitHub\footnote{\url{https://github.com/SageSELab/CodeSumStudy}} and Zenodo\footnote{\url{https://doi.org/10.5281/zenodo.4904024}} for replicating our experiments, including code and trained models, in addition to all of the data and examples used in our qualitative analysis of model errors.
\end{itemize}

\section{Background: Deep Learning for Code Summarization}
\label{sec:background}

This section outlines necessary background regarding our chosen evaluation dataset as well as the three neural code summarization models upon which we focus our empirical investigation.

\subsection{Dataset: Funcom}
In this study we make use of the Funcom dataset~\cite{funcomPaper}.\footnote{\url{http://leclair.tech/data/funcom/}} %~\cite{funcomDataset}. 
We selected this dataset primarily for three reasons: (i) this dataset was specifically curated for the task of code summarization, excluding methods more than 100 words and comments with $>$13 and $<$3 words or which were auto-generated, (ii) it is currently one of the largest datasets specifically tailored for code summarization, containing over 2.1M Java methods with paired JavaDoc comments, (iii) it targets Java, one of the most popular programming languages.\footnote{\url{https://octoverse.github.com}} In order to make for a feasible training procedure for our various model configurations, and to keep the dataset size in line with past work to which our studied models were applied (e.g., the size of the CodeXGlue dataset from \citet{codexgluePaper}, containing approximately 180000 Java methods and JavaDoc pairs, to which CodeBERT was applied) we chose to use the first 500,000 method-comment pairs from the \textit{filtered} Funcom dataset for our experiments. Note that we did not use the \textit{tokenized} version of the dataset as provided by \citet{funcomPaper} as each of our models has unique pre-processing constraints, described in detail in Appendix~\ref{app:preproc}.

\subsection{Models}
\begin{figure}[t]
\centering
\includegraphics[width=0.50\textwidth]{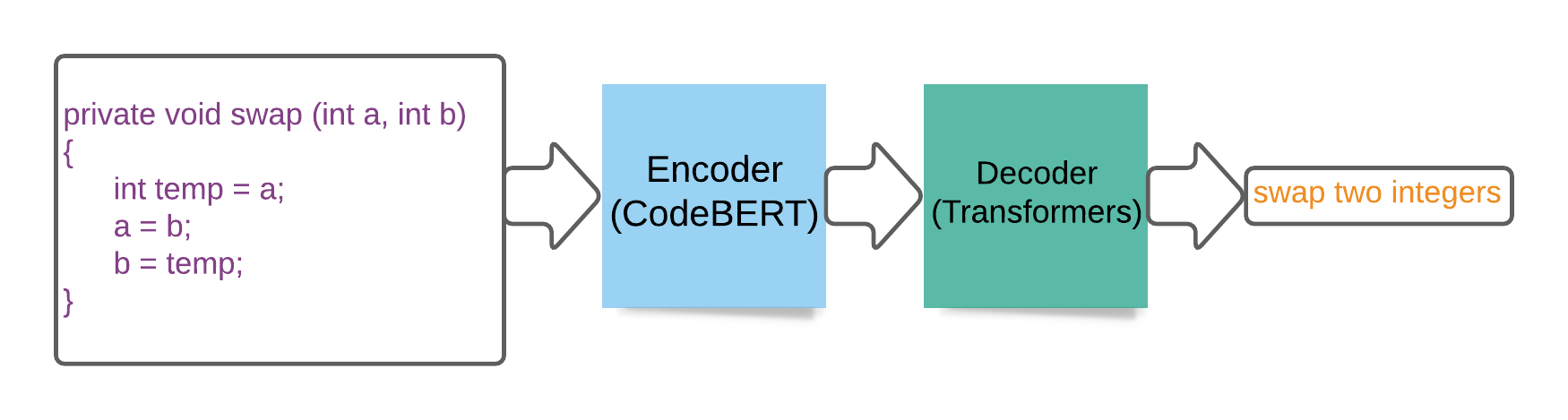}
\caption{Code to text translation using CodeBERT.}
\label{fig:codebert}
\end{figure}

\paragraph{CodeBERT}
CodeBERT~\cite{codebert} is a bimodal pre-trained model used in natural language (NL) and programming language (PL) tasks. This model supports six programming language tasks in various downstream NL-PL applications, e.g., code search, code summarization, etc. The architecture of the model is based on BERT~\cite{bert2019}, specifically following the RoBERTa-base~\cite{roberta} in using 125 million model parameters. The objectives of training CodeBERT are masked language modeling (MLM) and replaced token detection (RTD). Recently, Microsoft Research Asia introduced the CodeXGLUE benchmark that consists of 14 datasets for ten diversified code intelligence tasks~\cite{codexgluePaper}. They fine-tuned CodeBERT in code-to-natural-language generation tasks. CodeBERT was used as the encoder, with a six-layer self-attentive~\cite{attentionAll} decoder. An architecture for code-to-text translation using the CodeBERT encoder is shown in Figure~\ref{fig:codebert}. The dataset~\citet{codexgluePaper} used is derived from CodeSearchNet~\cite{codeSearchNet}.

\paragraph{NeuralCodeSum}
The second technique we study is NeuralCodeSum \cite{neuralCodeSum}. Here, the authors explored a transformer-based approach to perform the task of code summarization, using a self-attention mechanism to capture the long-term dependencies that are common in source code. In order to enable the model to both copy from already seen source code and to generate new words from its vocabulary, they employed a copy mechanism~\cite{See_2017}. One important distinction of source code that this model takes into account is that the absolute token position does not necessarily assist in the process of learning effective source code representations (i.e., \texttt{int a=b+c} and \texttt{int a=c+b;} both convey the same meaning). To mitigate this problem, they used the relative positioning of tokens to encode pairwise token relations. Additionally, the authors of this model also explored the integration of an abstract syntax tree (AST)-based source code representation. However, they found that the AST information did not result in a marked improvement in model accuracy.

\paragraph{code2seq}
The third model we consider in our study is code2seq \cite{code2Seq}, which is a widely utilized technique that was originally designed for the task of method name prediction. The authors of this work focused on capturing the true syntactic construction of source code by encoding AST paths. They showed that code snippets which exhibited differences in lines but that were designed for similar functionality often have similar patterns in their AST trees. To take advantage of this observation, code2seq uses an encoder-decoder architecture that attends to the constructed AST encoding to generate the resultant sequence. The authors experimented with Java method name generation as well as code captioning tasks. They compared their code captioning approach to CodeNN \cite{iyer_summarizing} using BLEU score, against which it illustrated improved performance.

\section{Design of the Empirical Evaluation}
\label{sec:eval}

To evaluate the performance of our three models applied to the task of code summarization, we perform both a \textit{quantitative} and \textit{qualitative} evaluation centered upon the following research questions:

\begin{itemize}[leftmargin=*]
    \item[] \textbf{RQ$\mathbf{_1}$}: \textit{How effective is each model in terms of predicting natural language summaries from Java methods?}\vspace{-0.5em}
    \item[] \textbf{RQ$\mathbf{_2}$}: \textit{What types of errors do our studied models make when compared to ground truth captions?}\vspace{-0.5em}
    \item[] \textbf{RQ$\mathbf{_3}$}: \textit{What differences (if any) are there between the errors made by different models?}
\end{itemize}

\subsection{Evaluation Methodology for RQ$\mathbf{_1}$}

In this subsection, we discuss how we split the dataset, the evaluation metrics we use, and how we configure our studied models for training.  

\subsubsection{Dataset Preparation and Metrics}
To adapt the Funcom dataset for our study, we first sampled the first 500k function-comment pairs from the \textit{filtered} Funcom dataset into training (80\%), validation (10\%) and testing (10\%) for our experiment, ensuring that the method-comment pairs between our training and testing datasets came from separate software projects (i.e., split by project), as suggested by the Funcom authors, in order to avoid artificial inflation of performance due to data snooping~\cite{funcomPaper}. As a comparison to past work, we illustrate the training, validation and test dataset sizes between the CodeXGLUE and Funcom datasets in Table~\ref{table:dataStatistics}. As mentioned earlier we preprocess the sampled dataset based on the requirements for each of our chosen models, and provide details in Appendix~\ref{app:preproc}. 

Prior work has explored the use of several reference-based metrics, e.g., BLEU, METEOR, and ROUGE-L for evaluating the performance of code summarization. In our study we make use of smoothed BLEU-4 as it was previously used to evaluate the CodeBERT model~\cite{codebert}. BLEU is the geometric average of $n$-gram precisions between the predicted and reference captions multiplied by a brevity penalty that penalizes the generation of short descriptions. We use the BLEU metric applying a smoothing technique ~\cite{lin2004orange}, which adds one count in the case of $n$-gram hits to address hypotheses shorter than $n$. In addition, we include METEOR~\cite{meteor} and ROUGE-L~\cite{rouge} in our study. METEOR computes the harmonic mean between precision and recall based on unigram matches between the prediction from a model and reference, also going beyond exact matches to include stemming, synonyms, and lemmatization. ROUGE-L computes the longest common subsequence-based F-measure between the hypotheses and references.

\begin{table}[t]
\centering
\begin{tabular}{c|ccc}
\toprule
     & \textbf{\footnotesize{Training}} & \textbf{\footnotesize{Dev}} & \textbf{\footnotesize{Testing}}\\
    \midrule
    CodeXGlue & 164923 & 5183 & 10955\\
    \midrule
    \textbf{Funcom} & \textbf{400000} & \textbf{50000} & \textbf{49997}\\
    \bottomrule
\end{tabular}
\caption{Data Statistics. We use the Funcom dataset.}
\label{table:dataStatistics}
\vspace{-1em}
\end{table}

\subsubsection{Model Configurations and Training}
We train, validate and test the three models described in \secref{sec:background} for the task of summarizing Java methods in natural language. A subset of model hyperparameters for all three studied deep learning models is shown in Table~\ref{table:hyperparameters}. We preprocess the dataset for each of the models according to their individual requirements and select the hyperparameters for each of the models based on the optimal settings from prior work.Additionally, we apply some global preprocessing that is common to all models, taken from recent work on language modeling for code~\cite{mastropaolo2021studying}. Initially, we remove all the comments that exist inside methods, as the commented code could lead to poor predictions. Next, all the JavaDoc comments are filtered keeping only the description of the method. Finally, we clean HTML and remove special characters from the JavaDoc captions. We provide a detailed account of our preprocessing and training techniques in Appendix~\ref{app:preproc} and in our publicly available resources.  

\paragraph{CodeBERT Model Configurations and Training:}
We use the open-source implementation\footnote{\url{https://github.com/microsoft/CodeXGLUE/tree/main/Code-Text/code-to-text}} made available by Microsoft to fine-tune CodeBERT using the Funcom dataset. We utilized the optimal model configurations for this model used to train on the CodeXGlue~\cite{codexgluePaper} dataset with hyperparamters tuned on the Funcom dataset.

\paragraph{NeuralCodeSum Model Configurations and Training:}
We use the open-source implementation of NeuralCodeSum\footnote{\url{https://github.com/wasiahmad/NeuralCodeSum}} to train the model in our study. We performed one additional preprocessing step than typical with this model, splitting camel-case words. The dropout rate is set to 0.2 and we train for a maximum of 1000 epochs. Additionally, we stop training if validation does not improve after 20 iterations.

\begin{table}[t]
\centering
\small
\begin{tabular}{@{}l|l|l|l@{}}
\toprule
    \textbf{Hyper-} & \textbf{CodeBERT} & \textbf{Neural-} & \textbf{code2seq}\\
    \textbf{parameters} & & CodeSum \\
    \midrule
    Batch Size & 16 & 64 & 512\\
    Beam Size & 16 & 4 & 0\\
    Optimizer & Adam & Adam & Momentum\\
    Learning Rate & 0.00005 & 0.0001 & 0.01\footnotesize{+decay} \\ 
    \#epochs & 15 & 38 & 39\\
    \bottomrule
\end{tabular}
\caption{Model Hyperparameters.}
\label{table:hyperparameters}
\end{table}

\paragraph{code2seq Model Configurations and Training:}
We make use of the publicly available implementation of code2seq.\footnote{\url{https://github.com/tech-srl/code2seq}} To use the Funcom dataset, we had to prepare the AST node representation using a modified dataset build script.\footnote{\url{https://github.com/LRNavin/AutoComments}} The original dataset build script was designed to predict the method name whereas we modify it to predict summaries. One problem we faced representing Funcom methods as ASTs is that there were some code examples which could not be parsed into an AST representation mainly because of the imposed minimum code length threshold and the method not having any AST-Paths. As a result, we were able to train code2seq on only a subset of the Funcom dataset ($40009/ 50000 \approx 80.02\%)$. To train the model we made use of large batch sizes (e.g., 256 and 512) as we noted smaller batch sizes resulted in instability. As code2seq was originally designed to predict method names, we also made some changes in the model parameters to facilitate longer prediction sequences, which we give in Appendix~\ref{app:hyperParametersAppendix}.

\subsection{Evaluation Methodology for RQ$\mathbf{_2}$ \& RQ$\mathbf{_3}$}
We performed a manual, qualitative analysis on the output of the three models\footnote{Some examples of the predictions are shown in Appendix~\ref{sec:case}} to answer \textbf{RQ$\mathbf{_2}$} and \textbf{RQ$\mathbf{_3}$} in order to better understand and compare the various types of errors each model makes. The methodology we follow to categorize the model prediction errors follows a procedure inspired by open coding~\cite{openCodingBook}, which has been used in prior studies to categorize large numbers of software project artifacts~\cite[\textit{inter alia}]{mutationTesting}. Initially, we randomly selected a small number of samples from our validation split of the Funcom dataset, and applied each of our three models to generate captions. The four annotators\footnote{All annotators are also authors of this study.} then met and discussed the samples to derive an initial set of labels that described deviations from the ground truth. We found that 15 methods (each with three predictions, one from each of our studied models) were enough to reach an initial agreement on the labels. Note that we use the ground truth captions as a ``gold set" in order to orient our analysis to a shared understanding among annotators and to limit potential subjectivity.

Next, we conducted two rounds of \textit{independent} labeling, wherein three annotators independently coded a samples of method-comment pairs and predicted comments, such that two annotators independently coded each sample. Here we define a ``sample'' as a method $\leftrightarrow$ gold-comment pair, \textit{and} the three resulting predictions from CodeBERT, NeuralCodeSum, and code2seq respectively for the method. During this process, annotators were free to add additional labels outside of the initial set if they deemed it necessary. The first round of labeling consisted of 148 samples in total, amounting to $148 \times 3 = 444$ predictions from our studied models. After the independent labeling process, the authors met to resolve the conflicts among the labels. This initial round of coding resulted in a disagreement on $\approx82\%$ of the samples wherein author discussion was needed in order to derive a common agreed upon label. There were two main reasons for this relatively high rate of disagreement: (i) the authors created some category labels with similar semantic meanings, but different labels, and (ii) some of the authors had different interpretations of shared meanings. However, through an extensive discussion, the conflicts were resolved and a shared understanding reached. The second round of independent labeling consisted of 50 samples, and resulted in a disagreement rate of only $\approx 27\%$, illustrating the stronger consensus among authors. We derive the taxonomy presented in~\secref{studyResults} from labels present after both rounds of our open coding procedure.

\begin{table*}[t]
\centering
%\small
\begin{tabular}{c|c|c|c}
\toprule
    \textbf{Models} & \textbf{Smoothed BLEU-4} & \textbf{METEOR} & \textbf{ROUGE-L}\\
    \midrule
    CodeBERT & \textbf{24.15} & \textbf{30.34} & \textbf{35.65}\\
    NeuralCodeSum & 21.50 & 27.78 & 33.71 \\
    code2seq & 18.61 & 27.31 & 33.52\\
    \bottomrule
\end{tabular}
\caption{Evaluation Results with three metrics. CodeBERT is consistently better than the other two models.}
\label{bleu4}
\end{table*}

\section{Evaluation Results}
\label{studyResults}

In this section, we will discuss the \textit{quantitative} and \textit{qualitative} results from our empirical study in order to answer our research questions.  

\subsection{RQ$\mathbf{_1}$ Results: Evaluation Based on Reference-Based Metrics}
To perform the evaluation on the Funcom dataset, we use the optimal hyper-parameters shown in Table~\ref{table:hyperparameters} for the three deep learning models. NeuralCodeSum could not predict natural language descriptions for some examples ($\approx80$). The most likely reason for this situation is the errors in processing code or docstring tokens. Table~\ref{bleu4} shows the quantitative results obtained based on smoothed BLEU-4, METEOR, and ROUGE-L scores. The results show that CodeBERT performs best among the three models. We believe that the reason we observe CodeBERT achieving this level of performance is that this model is pre-trained on both bimodal data and unimodal data (wherein bimodal data refers to the coupled code and natural language pairs and unimodal data refers to either natural language descriptions without code snippets or code snippets without natural language descriptions~\cite{codebert}).

\paragraph{Statistical significance}
In addition to calculating the evaluation scores (i.e. smoothed BLEU-4, METEOR, ROUGE), we conducted statistical significance tests for all three metrics to assess the validity of the obtained results. We took 19009 examples from the test dataset and used pairwise bootstrap re-sampling~\cite{koehn-2004-statistical} between all 3 model predictions. In comparison to NeuralCodeSum, we found CodeBERT performs better with a mean score increase (BLEU-4 2.8, METEOR 2.9, ROUGE 2.2) at a 95\% confidence interval, thus indicating a performance delta that is statistically significant.

\begin{figure*}[t]
\centering
%\vspace{-1em}
\includegraphics[width=\textwidth]{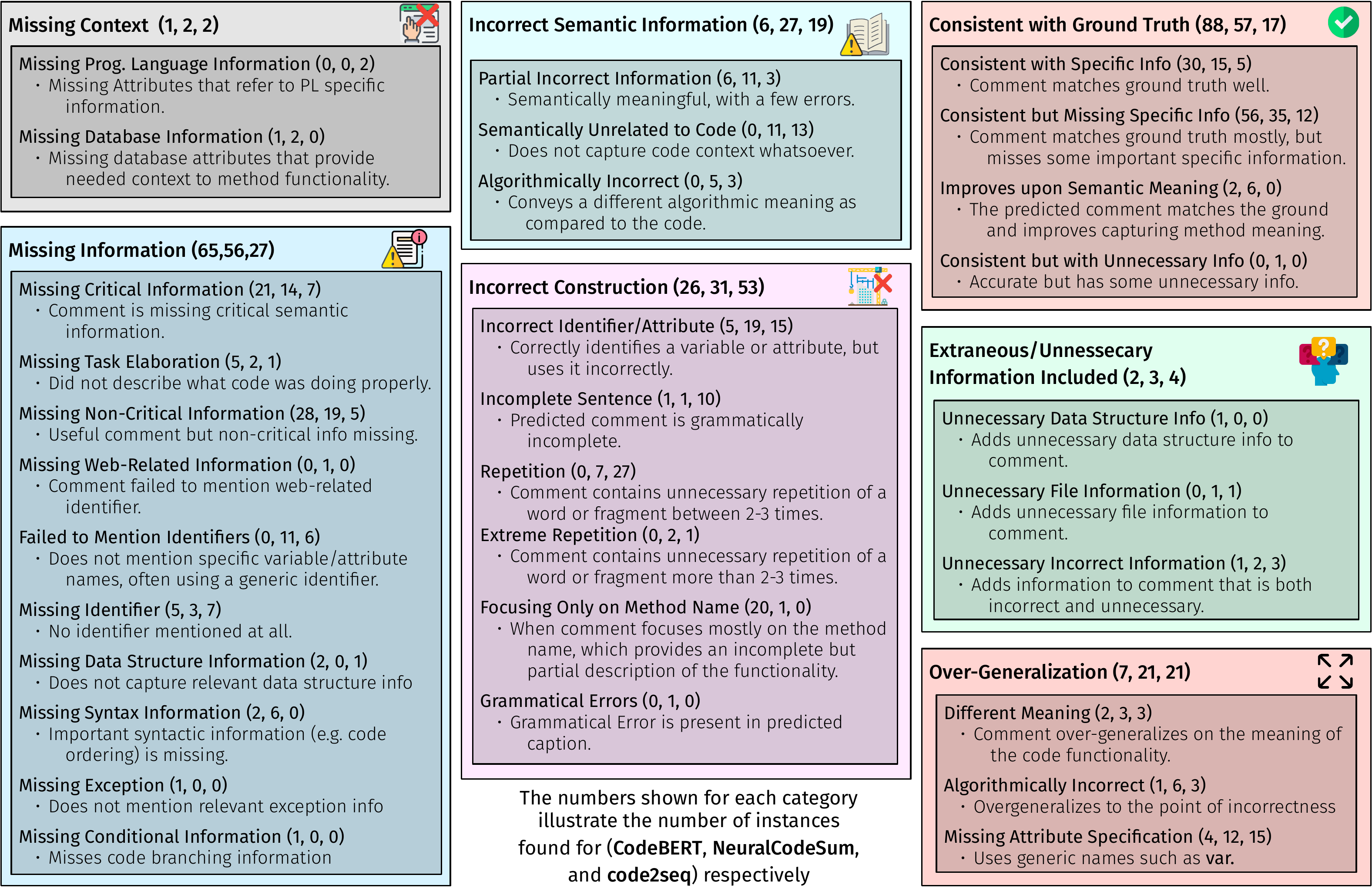}
\caption{Taxonomy of the Errors Between the Generated Summaries and the Ground Truth}
\vspace{-0.5em}
\label{fig:taxonomyErrors}
\end{figure*}

\subsection{RQ$\mathbf{_2}$ Results: Types of errors} 
In the first round of our study that included $148 \times 3 = 444$ samples, we were able to classify the errors for 398 generated natural language descriptions from the models from the validation dataset. The remaining 46 descriptions that were not classified as predictions were not made by the models due to errors in parsing and one error in processing code tokens. This singular error was due to the fact an entire code snippet was commented out, and our models do not process commented code. Thus, we did not include the predictions for the three different models for that code snippet in our study. In the other 43 cases, the code2seq model could not generate predictions because the model was not able to parse the AST.

Our error taxonomy derived after both rounds of the open coding process is shown in Figure~\ref{fig:taxonomyErrors}. The taxonomy consists of seven high-level categories with each consisting of multiple lower-level sub-categories. To elaborate, \texttt{\small \textbf{Semantically Unrelated to Code}} is a sub-category of \texttt{\small \textbf{Incorrect Semantic Information}}. Note that one category \texttt{\small \textbf{Consistent with Ground Truth}} is dedicated to those captions that generally \textit{matched} the ground truth, which we include for completeness.  The numbers that are shown beside the name of the sub-categories illustrate the number of errors for CodeBERT, NeuralCodeSum, and code2seq respectively. The numbers shown beside the categories' names represent the cumulative sum of the sub-categories. We provide a small number of examples of these categorizations in Appendix~\ref{sec:case}, and provide all labeled examples in our public resources on GitHub and Zenodo. We make the following notable observations resulting from our derived taxonomy:

\begin{itemize}[leftmargin=*, nolistsep]
    \item Encouragingly, among the samples studied, the largest category of samples did not display significant errors, falling into the \texttt{\small \textbf{Consistent with Ground Truth}} category ($162/535 \approx 30.28$\%). This category is the most frequent among all, but we do see CodeBERT (unsurprisingly) exhibit the largest number of reasonable summaries.

    \item The most prevalent error category exhibited among our studied models was that of \texttt{\small \textbf{Missing Information}} ($148/535 \approx 27.66$\%) followed by the \texttt{\small \textbf{Incorrect Construction}} category ($110/535 \approx 20.56$\%). This seems to indicate that one of the biggest struggles for current neural code summarization techniques is related to the inclusion of various types of necessary information in the summary itself, followed by issues in properly constructing comment syntax.
    
    \item The models also either incorrectly recognized or failed to recognize salient identifiers that were needed to understand method functionality in a non-negligible number of cases ($71/535 \approx 13.2\%$). This suggests that mechanisms for identifying \textit{focal identifiers} i.e., those that might prominently contribute to describing the functionality, could be beneficial, similar to past work on identifying focal methods~\cite{Abdallah:ICSM10}.
    
    \item Some of the models exhibited generated summaries that over-generalized to the detriment of the summary meaning ($49/535 \approx 9.15\%$) , whereas very few summaries contained extraneous information.
    
    \item Further study is needed to gain a better understanding of the various facets of the \textit{critical information} and \textit{non-critical} information that captions were missing. For instance, we plan to explore whether the necessary information is contained within the code itself, or perhaps in semantically related methods. We leave this for future work.
    
\end{itemize}

\subsection{RQ$\mathbf{_3}$ Results: Comparison of three different models}
One advantage of the formulation of our empirical study is that we are able to compare the various shortcomings of our studied models as they relate to our qualitative error analysis. To this end, we make the following notable observations:

\begin{itemize}[leftmargin=*]
    \item The most frequent error categories for CodeBERT and NeuralCodeSum are \texttt{\small \textbf{Consistent but Missing Specific Information}} (CodeBERT: $56/197 \approx 28.42$\% and NeuralCodeSum: $35/197 \approx 17.77$\%). However, for code2seq, the most frequent category is \texttt{\small \textbf{Repetition}} ($27/141 \approx 19.15$\%). 
    
    \item A non-negligible number of predictions from CodeBERT fall into the \texttt{\small \textbf{focusing Only on the Method Name}} category ($20/197 \approx 10.15$\%). This may suggest a reliance of the model on descriptive method names in order to produce reasonable summaries.
    
    \item NeuralCodeSum and code2seq produce a small number of predictions that are \texttt{\small \textbf{Semantically Unrelated to Code}}. However, we did not find any such cases for CodeBERT.
    
    \item Similar to our quantitative evaluation, we find that CodeBERT performs best, but suffers from a large number of errors related to \texttt{\small \textbf{Missing Information}}. In future work, we will investigate the adaptation of source coverage techniques~\cite{cohn-etal-2016-incorporating,mi-etal-2016-coverage} to our task to mitigate this issue.     
\end{itemize}

\section{Discussion \& Learned Lessons}

\noindent \textbf{Takeaway 1: The CodeBERT model illustrates improved performance on the Funcom dataset as compared to CodeXGLUE, likely due to the filtering steps undertaken in its construction.} Previously, the CodeBERT model was fine-tuned on the CodeXGlue dataset and the smoothed BLEU-4 score obtained on the Java dataset was 17.65~\cite{codexgluePaper}. However, we fine-tuned the model on the Funcom dataset and obtained a smoothed BLEU-4 score of 24.15. We believe there are two primary contributing factors to this observation: 1) A higher volume of data, and 2) filtering strategies. CodeXGLUE only provides 164923 training examples, whereas we used 400000 Java Methods and Javadoc pairs during he fine-tuning process. Moreover, The CodeXGLUE dataset is obtained from CodeSearchNet and the documents that contain special tokens (e.g., $<$img$>$ or https:) are filtered. In our preprocessing, we did not completely remove such data in the preprocessing; we only remove the HTML and special characters from the JavaDoc captions. We hypothesize that such characters may contain important information and as such lead to more effective predicted summaries.

\noindent \textbf{Takeaway 2: Models that rely on statically parsing source code can lead to high numbers of missing/incomplete predictions.}
The preprocessing for the code2seq model includes generating strings from the AST node representation of each method. Unfortunately, it is difficult (or impossible) to construct a suitable AST representation for methods that fall under a certain token length threshold. As a result, about 19.98\% of the original dataset could not be fed into the code2seq testing module, and for which we could not generate any prediction for these examples. 

\noindent \textbf{Takeaway 3: Some of the generated summaries provide a semantic meaning similar to the ground truth, despite exhibiting fewer $n$-gram matches.} Our studied models can generate summaries that contain relevant semantic information which can be useful for code comprehension despite not perfectly matching the ground truth. For instance, let's consider the following example ground truth for a Java method, \textit{``this method sets the text for the heading on the component"}. The generated summary from the CodeBERT model is 
\textit{``sets the heading caption"}. Comparing these two descriptions will not necessarily result in a high BLEU-4 score. This suggests that a modification to the evaluation procedure for these models may provide a more realistic characterization of model performance in practice. For instance, measuring BERTScore in addition to other metrics for evaluation~\cite{zhang2020bertscore}\footnote{\url{https://github.com/Tiiiger/bert_score}} may help to better capture semantic similarities compared to purely symbolic similarities.

\noindent \textbf{Takeaway 4: Future techniques for Neural Code Summarization should carefully consider techniques for mitigating potential errors related to \textit{Missing Information}, and \textit{Incorrect Construction} as these are the most prevalent error types observed in our taxonomy.} Our error taxonomy provides concrete indicators on where different types of models stand to gain performance in order to make them useful for downstream deployment. In particular, we suggest that future research focuses on rectifying \texttt{\small \textbf{Missing Critical Information}} and \texttt{\small \textbf{Missing Non-Critical Information}} rather than \texttt{\small \textbf{Grammatical Errors}} or \texttt{\small \textbf{Unnecessary File Information}}.

\noindent \textbf{Takeway 5: Future studies should explore the combination of AST traversal based and self-attention mechanism-based approaches to perform robust comment generation.} AST-based approach is useful to provide syntax level information and it follows the structural tree traversal method to capture the global information. At the same time, we can see this approach is prone to errors like \texttt{\small \textbf{Repetition}} and \texttt{\small \textbf{Semantically Unrelated to Code}}. On the other hand, a self-attention mechanism is useful to capture the local information. So a multi-modal approach where standard encoders can be utilized to combine both AST-based and attention-based approaches can be a viable direction to explore further.

\noindent \textbf{Takeway 6: Robust evaluation metric(s) should be developed that specifically focus on source code - natural language translation.} Source code is fundamentally different from the natural language from a number of perspectives. For instance, it exhibits less significant word order dependency, the significance of appropriate syntax naming and mentioning, etc. So a robust code to natural language translation evaluation metric should consider assessment from both local and global levels. Standard machine translation metrics like BLEU, METEOR, ROUGE do not fully cover these factors. As such, we encourage future work to study and develop new forms of automated metrics for assessing this special case of machine translation.

\section{Related Work}
\label{sec:related}
% \begin{itemize}
%     \item Code to Comment Translation Techniques
%     \item Empirical study on the predictions
% \end{itemize}

\subsection{Code to Comment Translation}
Source code summarization is a topic of great interest in software engineering research. The aim is to automate a portion of the software documentation process by automatically generating summaries of a given granularity for a source code snippet (e.g., methods) to save developer effort. Techniques have evolved from using more traditional Information Retrieval (IR) and machine learning methods to utilizing artificial neural networks.

One of the earliest deep-learning-based source code summarization techniques is that by~\citet{iyer_summarizing}. The authors used an attention-based neural network to generate NL summaries from source code. The approach was applied to the C\# programming language and SQL. %However, we concentrate on generating summaries based on Java. 
Given the strong syntax associated with programming languages, researchers have also experimented with utilizing AST information for source code summarization. \citet{deepCode} used an AST traversal method to generate summaries. Additionally, ~\citet{summariesSubroutines} utilized structural code information by encoding ASTs. %We also adapted an AST-based approach in our model~\cite{code2Seq}.
Our goal in this study is to provide an overview on the performance of a variety of techniques, both sequence based (i.e., CodeBERT, NeuralCodeSum), and structure-based (i.e., code2seq), in order to examine differences in quantitative and qualitative performance across different types of models.
Recently, a more complex retrieval-augmented mechanism was introduced that combines both retrieval and generation-based methods for code to comment translation~\cite{retrievalaugmented}. Finally,~\citet{projectLevel} recently proposed a method that uses a vectorized representation of source code files. We plan to explore additional techniques such as these in future work. 

\subsection{Empirical Studies of Code Summaries and Code Summarization}
Although many deep learning models are capable of generating summaries from source code, very few researchers have focused on evaluating the errors made by the models from a human perspective. During an early study on this topic, \citet{codeFragment} tried to understand whether code summaries achieved the same level of agreement from multiple human perspectives. \citet{compareComments} performed a comparison based on the similarities of the summaries generated by a newly proposed model which aimed at including \textit{context} in code summaries. %We do not concentrate on understanding if the models can generate predictions that are as good as comments written by humans. We focused on understanding the errors that exist between the natural language descriptions generated by the models and the ground truths. M
However, most recent work on code summarization models, e.g.,~\cite{GNNBased,projectLevel} depend on machine translation metrics to measure the performance of the code summarization task. However, a recent study showed a necessity of revised metrics for code summarization~\cite{recentStudy}.

Perhaps the most closely related study to ours is that conducted by \citet{Gros:ASE'20}. In this study, the authors question the validity of the formulation of code summarization as a machine translation task. In doing so, they apply code and natural language summarization models to several recently proposed code summarization datasets and one natural language dataset. They found differences between the natural language summarization and code summarization datasets that suggests marked semantic differences between the two task settings. Additionally, the authors carried out experiments which illustrate that reference-based metrics such as BLEU score may not be well suited for measuring the efficacy of code summarization tasks. Finally, the authors illustrate that IR techniques perform reasonably well at code summarization. While this study derives certain conclusions that are similar to those in our work (e.g., the need for better automated metrics) our study is differentiated by our manually derived fault taxonomy.

To the best of our knowledge, no other study has taken on a large-scale qualitative empirical study with the objective of categorizing and understanding errors between automatically generated and ground truth code summaries. Thus, we believe this is one of the first papers to take a step toward a grounded understanding of the errors made by neural code summarization techniques -- offering empirically validated insights into how future code summarization techniques might be improved. %In this paper, we find the errors based on an open-coding process. We find the errors from two authors' perspectives and categorize the errors. We also resolve the conflicts between the authors. A taxonomy is proposed based on the findings of the errors. 

\section{Conclusion \& Future Work}
In this work we perform both quantitative and qualitative evaluations of three popular neural code summarization techniques. Based on our quantitative analysis, we find that the CodeBERT model performs statistically significantly better than two other popular models (NeuralCodeSu, and code2seq) achieving a smoothed-BLEU-4 score of 24.15, a METEOR score of 30.34, and a ROUGE-L score of 35.65. Our qualitative analysis highlights some the most common errors made by our studied models and motivates follow-up work on improving specific model attributes. 

In the future, we aim to expand our analysis to additional retrieval-augmented summarization techniques and to expand the scope and depth of our neural code summarization model error taxonomy.

\bibliographystyle{acl_natbib}
\bibliography{ms.bib}

\clearpage
\newpage
\appendix
\section{Hyper-parameters}
\label{app:hyperParametersAppendix}
In Table \ref{hyperParametersTable}, we show the hyper-parameters that are used in our adapted models. Code2seq model could not be trained using batch size 64 or 128 because of the instability occurred from the longer comment length. Originally, this model was designed to predict the method name. So we trained the model using batch size 512 in our final experiment and it required 39 epochs to train the model.
\begin{table}[htb]
\centering
\small
\begin{tabular}{p{2cm}|p{1.4cm}|p{1.3cm}|p{1.6cm}}
\toprule
    \textbf{\footnotesize{Hyper-parameters}} & \textbf{\footnotesize{CodeBERT}} & \textbf{\footnotesize{Neural-CodeSum}} & \textbf{\footnotesize{Code2Seq}}\\
    \midrule
    Maximum Source Length & 256 & 150 & 200\\
    Batch Size & 16 & 64 & 512\\
    Beam Size & 16 & 4 & 0\\
    Optimizer & Adam & Adam & Momentum\\
    Learning Rate & 0.00005 & 0.0001 & 0.01+exp. decay \\ 
    \#epochs & 15 & 38 & 39\\
    Dropout rate & 0.1 & 0.2 & 0.25\\
    \#Attention heads & 12 & 8 & --\\
    Early stopping & True & True & True\\
    \#layers & 6 & 6 & --\\
    \bottomrule
\end{tabular}
\caption{Model Hyperparameters}
\label{hyperParametersTable}
\end{table}
\section{Data Prepossessing}
\label{app:preproc}
We had to perform several preprocessing steps to make the dataset ready for training. Among all the three models, we removed comments inside methods, removed tags, clean HTML, lowercasing characters, removing special characters. For the NeuralCodeSum model, we applied an additional sub-tokenization step. For code2seq, we needed to prepare the AST representation of the code snippets. To do this, we used a modified \texttt{JavaExtractor}\footnote{\url{https://github.com/LRNavin/AutoComments}} which locates the Java methods and put them in a file where each line is for one method. Subtokenization is performed in between to tokenize the CamelCase attributes (i.e. \texttt{["ArrayList"->["Array", "List"]]}). The original dataset build script was designed to put the method name in the prediction window. The modified one puts the comment instead of a method name. In Table \ref{tab:ast}, a Java code, comment and the equivalent one line dataset instance (AST representation) is presented. While performing this step, some methods could not be parsed as this AST representation mainly because of the minimum method length threshold required for the parsing. In total, we could transform 80.02\% of our training dataset on which we trained the code2seq model. All the steps used in preprocessing are shown in Table~\ref{preprocessingTable}.

\begin{table*}[htb]
    % \centering
    \small
    \begin{tabular}{l|p{12cm}}
    \toprule
         \textbf{Original Method} &  \texttt{public Type getType() \{ return m\_type; \} }
        \\\hline
          \textbf{Comment} & returns the type of this technical information 
         \\\hline
          \textbf{AST represtation} & \texttt{returns|the|type|of|this|technical|information type,Cls0|Mth|Nm1,get|type type,Cls0|Mth|Bk|Ret|Nm0,m|type get|type,Nm1|Mth|Bk|Ret|Nm0,m|type} \\
         \bottomrule
    \end{tabular}
    \caption{AST representation of java method for code2seq training}
    \label{tab:ast}
\end{table*}

\begin{table}[htb]
\centering
\small
\begin{tabular}{p{2.5cm}|p{1.5cm}|p{1.5cm}|p{1.5cm}}
\toprule
    \textbf{\footnotesize{Preprocessing}} & \textbf{\footnotesize{CodeBERT}} & \textbf{\footnotesize{Neural-CodeSum}} & \textbf{\footnotesize{Code2Seq}}\\
    \midrule
    removed comments inside methods & \checkmark  & \checkmark & \checkmark\\
    removed tags for comments and methods & \checkmark  & \checkmark & \checkmark\\
    HTML cleaning & \checkmark  & \checkmark & \checkmark\\
    Sub-tokenization & & \checkmark & \checkmark\\
    Lowercase & \checkmark & \checkmark & \checkmark\\
    removing special characters & \checkmark & \checkmark & \checkmark\\
    \bottomrule
\end{tabular}
\caption{Preprocessing}
\label{preprocessingTable}
\end{table}

\section{Case Study}
\label{sec:case}
In Table \ref{tab:case}, model predictions are given with the ground truth and assigned error categories.
% \begin{landscape}

\begin{sidewaystable*}[htb]
    \centering
    \small
    \begin{tabular}{p{5cm}|p{3.5cm}|p{4.5cm}|p{4.5cm}|p{4.5cm}}
    \toprule
    \multirow{2}{*}{\textbf{ Java Method}} & \multirow{2}{*}{\textbf{Human written comment}} & \multicolumn{3}{c}{\textbf{Prediction (Catagory)}} \\ 
     & & \textbf{CodeBert}  & \textbf{NeuralCodeSum} & \textbf{Code2Seq}  \\
    \midrule
    \texttt{public float getDashPhase() \{ return dashPhase;\}} & gets the dash phase of the basicstroke & gets the dashphase (Consistent but Missing Specific Info) & returns the phase of dealpoint (Partial Incorrect Information) & get the current velocity of the current value (Semantically Unrelated to Code)  \\\hline
    \texttt{public void setReadTimeout(int timeout) \{ if (0 > timeout) \{ this.readTimeout = timeout; \}} & sets the timeout value in milliseconds for reading from the input stream & sets the read timeout (Focusing Only on Mehod Name) & sets the number of milliseconds to wait for a response before timing out (Consistent with Specific Info) & sets the timeout to read from the server (Consistent but Missing Specific Info)  \\\hline
    \texttt{public String numRulesTipText() \{ return "Number of rules to find."; \}} & returns the tip text for this property & gets the numrulestiptext attribute of the appletlocale object (Unnecessary Incorrect Information) & returns the number of labels in the rule set (Semantically Unrelated to Code) & returns the text of the text of the current text (Repetition)  \\\hline
    \texttt{public Items withTotalResults(BigInteger value) \{ setTotalResults(value); return this;\}} & sets the value of the totalresults property	& sets the total results (Consistent with Specific Info) &	sets the total number of items in the group (Incorrect Identifier/Attribute) &	returns the total results for the given var (Missing Critical Information) \\\hline
    \texttt{public String getSchema() \{ return fSchema; \}} & returns a path to the xml schema of a extension point & returns the schema (Consistent but Missing Specific Info) & returns the name of the xml schema (Missing Non-Critical Information)	& returns the schema schema (Repetition)
    \\\hline
    \texttt{public boolean isCreateIds() \{ return createIds; \}} & returns true if the model automatically creates ids and resolves id collisions & returns the createids (Partial Incorrect Information) & returns the \} (Missing Syntax Information)	& returns whether the is the default id (Missing Attribute Specification)
    \\\hline
    \texttt{public PromotionEligibilityRequir
    ement withQuantity(Integer value) \{setQuantity(value); return this; \}} & sets the value of the quantity property & returns a quantity (Different Meaning) & sets the value of this property any previous animations are stopped (Unnecessary Incorrect Information)	& sets the var (Missing Attribute Specification)
    \\\hline
    \texttt{public void setHeight(int height) \{ \_containerHeight = height; \}} & this method sets the minimum height of the table in pixels & sets the height of the container (Consistent but Missing Specific Info) & sets the height of the image (Incorrect Identifier/Attribute)	& sets the height of the image (Incorrect Identifier/Attribute)
    \\\hline
    \texttt{public void testForConnection() throws Exception \{ serverControl.ping(); \}} & try to test for a connection throws exception if unable to get a connection & test for a connection (Consistent but Missing Specific Info) & tests the connection to the server (Consistent but Missing Specific Info)	& test for the getter of the property (Missing Attribute Specification)
    \\\hline
    \texttt{public void setPath(Path path) \{ mPath = path.toString(); \}} & sets the value of the "path" attribute & sets the path (Consistent but Missing Specific Info) & sets the path (Consistent but Missing Specific Info)	& sets the path to the path of the path are not relative to the path of the path (Extreme Repetition)
    \\
    
    \bottomrule
    \end{tabular}
    \caption{Detailed case study of model predictions with ground truth}
    \label{tab:case}
\end{sidewaystable*}
% \end{landscape}
% \begin{sidewaystable}
%     \centering
% \caption{Wide table}
%     \label{tab:wide-item-tbl}
% \begin{tabularx}{\textwidth}{|*{4}{>{\RaggedRight\arraybackslash}X|}}
%     \hline
% a  &  b  &   c  &   d  \\
%     \hline
% \end{tabularx}
% \end{sidewaystable}

\end{document}